\documentstyle[12pt,psfig]{article}
\begin{document}
\begin{flushright}
CU-TP-875
\end{flushright}
\begin{center}
{\Large\bf
Equation of state and collision rate\\ 
tests of parton cascade models$^{\star}$
}\\[2ex]
{Bin Zhang$^{a}$, Miklos Gyulassy$^{a}$, and Yang Pang$^{a, b}$\\
{$^{a}$\em Physics Department, Columbia University,}\\
{\em New York, NY 10027, USA}\\
{$^{b}$\em Brookhaven National Laboratory,}\\
{\em Upton, New York, 11973, USA}}
\\[2ex]
\end{center}
%\title{
%Equation of state and collision rate\\ 
%tests of parton cascade models
%}
%\author{Bin Zhang$^{a}$, Miklos Gyulassy$^{a}$, and Yang Pang$^{a, b}$\\
%{$^{a}$\em Physics Department, Columbia University,}\\
%{\em New York, NY 10027, USA}\\
%{$^{b}$\em Brookhaven National Laboratory,}\\
%{\em Upton, New York, 11973, USA}}
%\date{}
%\maketitle

\begin{abstract}
\noindent We develop two further numerical tests of 
parton cascade models: the ideal gas equation of state and the collision
frequency tests. The equation of state test checks 
the initial momentum distribution generator and free expansion
dynamics in periodic inhomogeneous geometries. 
The collision rate test is sensitive to spatial inhomogeneities and the
collision algorithm. These tests are applied to the recently
developed  ZPC parton cascade model. The tests  helped uncover
unphysical  correlations induced by one of the commonly used  random number 
generators and showed the necessity of particle subdivisions
for convergence to the exact analytic limit.
\end{abstract}
\vspace{5cm}
\begin{picture}(500,5)(0,0)
\put(1,2){\line(1,0){180}}
\end{picture}

{\footnotesize
$^\star$This research was supported by the U.S. Department of Energy under
contract No. DE-FG02-93ER40764. Y. Pang was partially supported by the
U.S. Department of Energy grant DE-FG02-92ER40699, DE-AC02-76CH00016, and
Alfred P. Sloan Foundation.}\\[2ex]

\newpage

\section{Introduction}

Many observables in nuclear collisions are difficult to calculate analytically
because the number of particles is neither large enough to justify rigorously
the application of statistical mechanics 
 nor small enough to justify impulse approximations. In most cases, 
numerical simulations of transport equations
are required to  compare theory with experiment. However,  due to the
complexity of the algorithms employed and the number of untested 
dynamical assumptions 
\cite{ypang1,kwerner1,kkgeiger1,bzhang1}, it is not straight forward
to check or even reproduce the numerical results from parton cascade
codes. Recently, several analytic tests were proposed to help
check  the accuracy  and identify limitations of
such codes. In the context of earlier  non-relativistic
transport models tests such as comparing with 
 the analytic Krook-Wu model \cite{gwelke1} have been useful.
In the context of hydrodynamics,
 numerical hydrodynamic codes have been tested in cases
of expansion of baryon free matter into vacuum 
\cite{drischke1} and ``slab-on-slab" 
collision \cite{drischke2}.

To address the new generation of parton cascade,
a new Open Standard for Codes and
Routines (OSCAR)\cite{oscar1}  has been developed
to  enable objective testing of essential components of algorithms and
ensure reproducibility of numerical results.
In this context, we have proposed in ref.\cite{mgyulassy1}
 a dynamical test of the evolution of the transverse energy in
$1+1$ dimensional expansion for cascade models. Another test 
of the frame and scattering scheme dependence of cascade models 
was proposed in \cite{gkortemeyer1,bzhang2}. 
In this paper, we propose two further tests for the OSCAR 
standard: the equation of state and the collision rate test of 
one-component relativistic classical gases. 

The tests provide  information about the nature of the code's
initial momentum distribution and check its evolution
algorithm in free expansion in periodic  spatial
inhomogeneities against nontrivial analytic expressions.
The transverse to parallel momentum flux ratio
tests the evolution of space-time and momentum space correlations.
The asymptotic homogeneous equilibrated state tests
the equation of state of the model against the expected
ideal gas laws. The collision frequency test checks
the basic  collision algorithm. 

We apply these tests to the  newly developed
ZPC parton cascade model \cite{bzhang1}.
In order to  pass these tests, 
 we uncovered and fixed a
problem with one of the random number generators
used in ZPC. We discuss that example in detail
to illustrate the importance of applying such numerical
tests. We propose to add these tests 
to the group already in the OSCAR standard\cite{oscar1}.

The paper is organized as follows: In section~2, we recall basic statistical
mechanics relations of ideal gas and calculate the free streaming 
evolution of periodic slab initial spatial distribution.
In addition the collision rate is discussed. 
Numerical tests of ZPC are compared to
analytical predictions in section~3.  We conclude by emphasizing the
importance of testing cascade codes.

\section{Analytic Tests}
\subsection{Equation of State }

The partition function for an ideal relativistic classical
gas with mass $m$ and degeneracy $\gamma$
is given by \cite{stocker} ($\hbar=c=k_{\mbox{\scriptsize{B}}}=1$)
\begin{equation}
Z(N,T,V)=\frac{1}{N!}\left( \frac{\gamma V}{2\pi^2}(T m^2) K_2(m/T)
e^{m/T}
\right)^N
\;\; , \end{equation}
In the $N\gg 1$ limit, the free energy, $F=-T\log Z$  is well approximated by
\begin{equation}
F(N,T,V)=-N \left( T \log(\frac{\gamma}{2\pi^2}\frac{V T m^2}{N} 
K_2(m/T))
+m+T \right) 
\;\; . \end{equation}
The pressure, $P=-\partial F/\partial V$  is given by the well known
ideal gas law
\begin{equation}
P(N,T,V)= \rho T
\;\; , \end{equation}
in terms of the density $\rho=N/V$, and  the energy density
is given by
\begin{equation}
\epsilon(N,T,V)=
\rho m \left( K_1(\beta m)/K_2(\beta m) + 3T/m \right)
\;\; . \end{equation}
where $\beta=1/T$.
For fixed $N,V$ an interesting quantity that reflects the softness of
the equation of state is
\begin{equation}
\frac{P}{\epsilon} =\frac{1}{3 +
\beta m K_1(\beta m)/K_2(\beta m)}
\;\; . \end{equation}
In the $m/T\ll 1$ relativistic limit, $P/\epsilon\approx 1/3$,
while in the nonrelativistic limit, $P/\epsilon\approx T/m$,

In a cascade simulation, these thermodynamic quantities can be measured
as a function of time by
computing the spatially averaged energy-momentum tensor 
\begin{equation}
<T^{\mu\nu}>=\frac{1}{V}\sum_i\frac{p_i^\mu p_i^\nu}{E_i}.
\label{em1}
\end{equation}
Then $\epsilon=\langle T^{00}\rangle$ and $P=\langle T^{11}\rangle$. 
In section~3, we will compare the analytic results of the pressure-energy
density ratio as a function of $m/T$, and the energy density as a function of
$T$. 

\subsection{Free Expansion In Slab Geometries}

A simple test of equilibration in a {\em periodic} box 
of volume $V=L^3$ and fixed
particle number is provided by taking the initial spatial distribution 
to be confined to one half of the box (say with $x>0$):
\begin{equation}
f(x,p,0) =\rho_0 (1 +  \frac{4}{\pi} \mbox{Im}  
\tanh^{-1}[\mbox{e}^{2
\pi \mbox{\scriptsize{i}} x/L}]) f(p).
\label{f0}
\end{equation}
In the above, $\rho_0 = N / V$ is the equilibrium particle density, $L$ is the
length of the box, and 
\begin{equation}
f(p)= \frac{\exp(-\beta \sqrt{m^2 +p^2})}{4\pi (m^2/\beta) K_2(\beta m)}.
\end{equation} 
For the free streaming case, $f(\vec{x},\vec{p},t)=f(\vec{x}-\vec{p}
t/p^0,\vec{p},0)$.  

For the interesting case of massless partons, the momentum flux component,
$T^{11}$ evolves from its initial value
\begin{equation}
T^{11}(x,0)= T^{11}(\infty)(1 +  \frac{4}{\pi} \mbox{Im} 
\tanh^{-1} [ \mbox{e}^{2 \pi \mbox{\scriptsize{i}} x/L}))] 
\label{txx0}
\end{equation}
to its final value $ T^{11}(\infty)=\rho_0 \int \mbox{d}^3p \; p_x^2/p^0 f(p)$
via
\begin{equation}
\frac{T^{11}(x,t)}{T^{11}(\infty)}= 1 +
 \frac{a}{t^3}\sum_{l=0}^\infty
\frac{\sin (k_nx)}{n^4} (((k_nt)^2-2)\sin(k_nt)+2\,k_nt\cos(k_nt)).
\label{txxt}
\end{equation}
Here
$a= 3L^3/(2\pi^4)$, $n=2\,l+1$, and
$k_n=2\pi n/L$. 
The infinite sum can be expressed in terms of the Lerch $\Phi(z,n,a)$
functions, but in practice the series converges very rapidly.
For example the evolution at the ``midway'' points
$x=(2 k+1) L/4$ involves  damped oscillations with maxima and
minima occuring at times $t=(2 m +1)L/4$. 
At $x=L/4$, the ``$11$" momentum flux oscillates within the envelops
\begin{equation}
T^{11}(x=L/4,0)= T^{11}(\infty)\left(1 \pm 
\frac{\epsilon}{4} \left(\frac{3\,L}{t} -\frac{L^3}{8 t^3}\right) \right)
\label{env}
\end{equation}
reaching the envelop at times $t=(2 m +1)L/4$.
Note that free streaming in this geometry
leads to global homogeneous equilibrium
through mixing from neighboring slabs.
The deviation of the parton cascade evolution
from this result when collisions
terms are turned on 
are of physical interest as tests of collective
hydrodynamic behavior.

Another interesting probe of the evolution
is the transverse-longitudinal momentum flux anisotropy, 
\begin{equation}
A^{zx}(x, t) \equiv \langle T^{33}(x,t)\rangle/
\langle T^{11}(x,t)\rangle
\end{equation}
Initially in the $-L/2<x<0$ region we set $A^{zx}=0$
while in the $0<x<L/2$ region $A^{zx}=1$.
%with averages restricted to the $x<0$ region
%hould generally rise toward unity.
In the case of free streaming, this anisotropy is given by
\begin{equation}
A^{zx}(x, t)=\frac{1 +
 \frac{a}{t^3}\sum_{l=0}^\infty
\frac{\sin (k_nx)}{n^4} (\sin(k_nt)-k_nt\cos(k_nt))}
{1 +
 \frac{a}{t^3}\sum_{l=0}^\infty
\frac{\sin (k_nx)}{n^4} (((k_nt)^2-2)\sin(k_nt)+2\,k_nt\cos(k_nt))}.
\end{equation}

We note that of course
other non-equilibrium initial conditions, e.g., homogeneous spatial
distribution with anisotropic momentum distribution, can be set up to
test  dynamical relaxation toward global equilibrium.
In this paper we will study only the evolution
of spatial inhomogeneous conditions.

\subsection{Collision rate for an isotropic, homogeneous gas}

The collision rate per unit volume $W$ is another quantity that can be easily
monitored. It is related to 
the number of collisions $N_{\mbox{\scriptsize{c}}}$ in a period of time $t$
via: 
\begin{equation}
W=\frac{N_{\mbox{\scriptsize{c}}}}{Vt}.
\label{rate1}
\end{equation}

We take the integrated parton elastic cross section to be:
\begin{equation}
\sigma_{gg\rightarrow gg}=\frac{\pi}{\mu^2}.
\end{equation}

The total collision rate per
unit volume can be calculated from:
\begin{eqnarray}
W&=&<\sigma\rho_1\rho_2|\vec{v}_1-\vec{v}_2|> \nonumber \\
&=&\frac{1}{2}\frac{\pi}{\mu^2}\gamma^2
\int\frac{\mbox{d}^3p_1}{(2\pi)^3}\frac{\mbox{d}^3p_2}{(2\pi)^3}
\mbox{e}^{-\beta(E_1+E_2)}|\vec{v}_1-\vec{v}_2|,
\label{rate2}
\end{eqnarray}
and it is expected to be independent of the microscopic form of the
differential cross section. The $1/2$ on the right hand side of equation
(\ref{rate2}) takes into account the two identical incoming particles.
The identity of two incoming gluons has been taken into account in the
scattering cross section.
In appendix~A, we show that the above 6-dimensional integral can be reduced
to a 1-dimensional integral:
\begin{equation}
W=\frac{8T^6}{\pi^3\mu^2}F(\frac{2m}{T}),
\label{rate3}
\end{equation}
in which $F(x)=\int_x^\infty\mbox{d}y\;y^2(y^2-x^2)K_1(y)$.

\section{Numerical Tests of the ZPC model}

For the cascade calculations, we take the temperature $T=0.5,\;1,\;1.5$ GeV,
and set the total number of particles in the periodic box to be $N=4000$. The
volume is related to the density of particle at zero-chemical potential,
\begin{equation}
\rho=\frac{\gamma}{(2 \pi)^3} 4\pi\int_0^\infty
\mbox{d}p\;p^2\mbox{e}^{-\beta E},
\end{equation}
via $V=N/\rho$. In ZPC, there are $1000$ computation cells of volume
$V/1000$. We specify the cell size to the cascade code.

To study the mass dependence, we choose $3$ different masses: $0$, $2.5$, and
$5$ GeV. For the dependence on the scattering cross section, the interaction
length $\sqrt{\sigma/\pi}=1/\mu$ is set to be $0.5\lambda$, $\lambda$, and
$1.5\lambda$. In the 
above $\lambda$ is a rough estimate for the mean free path,
\begin{equation}
\lambda=\frac{1}{\rho\sigma}.
\end{equation}
We specify the screening mass $\mu$ to the cascade code.

The parameters for the cascade measurement are shown in 
Table\ (\ref{para1}). 

Fig.~1 shows the pressure-energy density ratio $P/\epsilon$ as a function of
$m/T$, and Fig.~2 shows the energy density as a function of temperature $T$ for
different particle masses. We see very good agreement between the predictions
and the cascade results. This indicates that the initial momentum distributions
are  correctly generated. There is no time dependence of the pressure and
energy density over a time period of $6$ fm during which each particle has
experienced $\sim 10$ collisions on average.

For periodic slab initial conditions, $T^{11}$ evolution at position $L/4$ in
the free streaming case is compared with prediction in Fig.~3. Also shown in
Fig.~3 is the result (filled circles) with interactions turned on. 
We see that interactions
reduce the rate of collective expansion compared to free streaming and also the
oscillations damp faster than in the free streaming case. For the
non-interacting free streaming case, good agreement of
prediction (dashed line) and cascade results (pluses) is shown for the $T^{11}$ evolution at $L/4$ in
Fig.~3 and time evolution of free streaming $T^{11}$ spatial distribution in
Fig.~4. Time evolution of free streaming $A^{zx}$ spatial distribution 
(Fig.~5) also agrees well with the prediction.

Fig.~6, 7 give scaled collision rate per unit volume as a function of $m/T$. 
The data points with the same $m/T$, and same $\mu$ overlap. The three data
sets at the same $m/T$ correspond to three different screening masses
$\mu$. The smaller the screening mass, the lower the collision rate. 

The collision rate is the same for the ZPC default Yukawa type 
scattering differential cross section and the straight line propagation. This
indicates the collision number loss is not due to particle shielding.

With small screening mass, the interaction range is large. When the
interaction range is much larger than the mean free path, non-causal
collisions become more abundant. To process the non-causal collision, we pick
up one collision out of several collisions for one particle according to the
ordering time (see ref.~\cite{bzhang2} for details). This process neglects
other collisions in the same non-causal collision set. Some of them will not be
recovered later.  The larger the interaction range, the larger percentage of
non-causal collisions. Hence, more collisions are neglected and the collision
rate is lower than expected.

In the dilute limit, the percentage of non-causal collisions out of total
number of collisions for massless
particles is proportional to the number of particles inside the causal sphere 
in the two colliding particle center of mass frame. The radius of the sphere is
proportional to the impact parameter. So 
\[\frac{N_{\mbox{\scriptsize{nonc}}}}{N_{\mbox{\scriptsize{total}}}} \propto 
\frac{1}{\sigma}\int_0^{\sqrt{\frac{\sigma}{\pi}}} 2\pi b\mbox{d}b
\frac{4\pi}{3}b^3\bar{\gamma}\rho=
\frac{8}{15}\bar{\gamma}\rho\frac{\sigma^{3/2}}{\sqrt{\pi}}.\]
$2\pi b\mbox{d}b/\sigma$
is the probability of having impact parameter $b$. Here, $\bar{\gamma}$ is 
for the averaged boost factor from lab frame to the two colliding particle
center of mass frame. In the case of massless particles, $\bar{\gamma}\approx
2$. $4\pi b^3\bar{\gamma}\rho/3$ gives the number of
particles in the sphere with radius $b$ in the two colliding particle center of
mass frame. The exact radius of the sphere depends on the definition of
non-causal collisions \cite{ypang1,bzhang2} and the collision prescription of
cascade.

The non-causal to total ratio is closely related to the ratio of interaction
length to the mean free path:
\[\chi=\frac{\sqrt{\frac{\sigma}{\pi}}}{\frac{1}{\rho\sigma}}=
\rho\frac{\sigma^{3/2}}{\sqrt{\pi}}.\]

We see $N_{\mbox{\scriptsize{nonc}}}/N_{\mbox{\scriptsize{total}}}$ decreases
linearly with $\chi$ as $\chi\rightarrow 0$. This motivates the algorithm of
reducing the non-causal collision percentage \cite{ypang1} by subdividing the
particles by a factor $l$ so that $\rho\rightarrow l\rho$ while decreasing
$\sigma\rightarrow \sigma/l$. This preserves the mean free path $1/\sigma\rho$
while $\chi\propto 1/\sqrt{l}\rightarrow 0$. 

The number of non-causal collisions, and total number of collisions for $4000$
particles with $m=0$, $T=0.5$ GeV, and a time period of $6$ fm is summarized in
table\ (\ref{para2}). In the table, $\mu$ is screening mass. $a$ gives the 
ratio of interaction 
length to the estimate of mean free path before rescaling. $l$ is the scaling
parameter, i.e., the total number of particles is increased by a factor of $l$
and the cross section is decreased by a factor of $l$ (but the number of
collisions is still for $4000$ particles). $\chi_1$ is the ratio of interaction
length to the estimate of mean free path including the rescaling, i.e.,
$\chi_1=a/\sqrt{l}$. $\chi_2$ is the ratio of interaction length to the
measured mean free path. The mean free path is measured through the formula,
$\rho/(2W)$, in which $W$ is the collision rate per unit volume and the
particles are moving at the speed of light. The ratio of the number of
non-causal collisions to the total number of collisions is also plotted in
Fig.~8 against the ratio of interaction length to the mean free path. In
Fig.~8, the open circles are data against $\chi_1$ and the filled circles are
against $\chi_2$. It shows that when the density increases, i.e., when $\chi$
increases, the difference between $\chi_2$ and $\chi_1$ increases. This tells
us that when density is high, the naive formula for the estimate of mean free
path, $\lambda = 1/(\rho\sigma)$, needs to be corrected. Also we see the data
deviate from linear formula and has a tendency of saturation when density is
large. This is consistent with the fact that the ratio should always be less
than $50\%$ from the definition of the non-causal collision.

When we fix $\mu$ and increase $l$ from $1$ to $5$, 
the total number of collisions goes up from $64700$ for $l=1$ to
$75400$ for $l=5$. For $l=10$ , the total number of collisions is $77800$. 
We see clearly the trend toward a constant value of total number of collisions
when $l$ is increased. The collision rate with $l=10$ is
shown in Fig.~6 as an open triangle (the $l=1$ data is shown as an open
circle). The collision rate with $l=10$ is within $1\%$ of the
analytic prediction. 

During the preliminary study of the collision rate, we found when $m/T=10$, the
collision rate is higher than the predicted rate. By looking more carefully
into the code, we found out that it was caused by the larger than statistical
fluctuations in the position distribution. When the fugacity $\lambda$ 
is not uniform in space, the collision rate per unit volume:
\begin{equation}
W=\int\frac{\mbox{d}^3x}{V}\lambda^2(x)I,
\end{equation}
in which:
\begin{equation}
I=\frac{\pi}{\mu^2}\gamma^2
\int\frac{\mbox{d}^3p_1}{(2\pi)^3}\frac{\mbox{d}^3p_2}{(2\pi)^3}
\mbox{e}^{-\beta(E_1+E_2)}|\vec{v}_1-\vec{v}_2|.
\end{equation}
By using the inequality,
\begin{equation}
\int\frac{\mbox{d}^3x}{V}\lambda^2(x)\ge 
\left(\int\frac{\mbox{d}^3x}{V}\lambda(x)\right)^2,
\end{equation}
and the fact that the system we prepared has zero chemical potential and
hence averaged fugacity is one, we arrive at:
\begin{equation}
W\ge W_0=
\frac{\pi}{\mu^2}\gamma^2
\int\frac{\mbox{d}^3p_1}{(2\pi)^3}\frac{\mbox{d}^3p_2}{(2\pi)^3}
\mbox{e}^{-\beta(E_1+E_2)}|\vec{v}_1-\vec{v}_2|.
\end{equation}
The equality holds only when the fugacity has no spatial dependence. 
This shows that when there are space clusters existing for some time
period, the collision rate is higher than that expected for the uniform system.

It was found only in $m/T=10$ and
not in other cases because in the $m/T=10$ case, particles are moving very
slowly, and they stay in clusters for much longer time.

We traced the origin of the non-uniform distribution. It was caused by some
correlation of random number generators (see Appendix~B). When we use ran1 from
\cite{press1}, and generate first the momenta for all the particles and then
generate the positions for all the particles, there are no abnormal
fluctuations. When we generate momentum and position together, we found
abnormal fluctuations. This does not occur when ran3 from \cite{press1} is
used. We correct the generation by separating the generation of particle
momentum and particle position.

\section{Conclusions}

From the above study, we show that the equation of state and the collision
rate can be used to test the initial conditions and collision mechanisms of
relativistic parton cascade. For massless particles, when the interaction range
is much larger than the mean free path, the cascade collision rate is lower
than the theoretical value. Other methods, e.g., particle partition, have to be
used to correct the collision rate. 

The comparison of free streaming and interacting cascade approach to
equilibrium indicate qualitative similarities of the two cases. However, the
damping and speed of collective motion are quite different. A detailed
comparison of free streaming, ideal hydrodynamics, and cascade approach to
global equilibrium in the case of half filled periodic box initial conditions
will be addresses in another paper \cite{mgyulassy2}.

As discussed in this paper, spatial distribution with larger than statistical
fluctuations gives higher than thermal reaction rate. HIJING \cite{xnwang1}
predicts initial spatial clusters of partons for nucleus-nucleus collisions at
collider energies. This implies higher than thermal collision rates
\cite{mgyulassy3} and many other interesting physical phenomena beyond the
widely used hot gluon scenario \cite{eshuryak1} predictions.

We emphasize the importance of using analytic tests in debugging
numerical simulation codes. The spatial distribution with abnormal fluctuations
illustrates well the usefulness of the analytic collision rate test. As more
components are added to the cascade code, more tests will be needed to
ensure the consistency of different parts of the cascade code and to enable
disentangling of the actual physical assumptions that define the model.\\[3ex]

{\Large \bf Acknowledgments}\\[1ex]

We thank S. A. Chin, P. Danielewicz, V. Koch, B. A. Li, S. Pratt, J. Randrup
for useful discussions. We also thank Brookhaven National Laboratory and
Lawrence Berkeley Laboratory for providing computing facilities. 

\appendix

\section{Reduction of the phase space integral in the collision rate
calculation} 

To calculate the collision rate per unit volume, equation\ (\ref{rate2}), we
first replace $|\vec{v}_1-\vec{v}_2|$ by 
\[\frac{\sqrt{s(s-4m^2)}}{2E_1E_2},\]
and get:
\begin{eqnarray}
W&=&\frac{1}{2}\frac{2\pi}{\mu^2}\gamma^2\int_{4m^2}^\infty\mbox{d}s\;
\sqrt{s(s-4m^2)}\int\frac{\mbox{d}^3p_1}{(2\pi)^32E_1}
\frac{\mbox{d}^3p_2}{(2\pi)^32E_2} \nonumber \\
&&\mbox{e}^{-\beta(E_1+E_2)}\delta(s-(p_1+p_2)^2).
\end{eqnarray}

The delta function can be used to integrate out one angle by using:
\begin{equation}
\delta(s-(p_1+p_2)^2)=\frac{1}{2p_1p_2}\delta\left(\cos \theta_2+
\frac{s-2m^2-2E_1E_2}{2p_1p_2}\right).
\end{equation}

After  carrying out the angular integrals, the collision rate becomes
\begin{eqnarray}
W&=&\frac{\pi\gamma^2}{4(2\pi)^6\mu^2}\int_{4m^2}^\infty
\mbox{d}s\;\sqrt{s(s-4m^2)}
\frac{4\pi p_1^2\;\mbox{d}p_1\;2\pi p_2^2\;\mbox{d}p_2}{E_1E_2\;2p_1p_2} 
\nonumber \\
&&\mbox{e}^{-\beta(E_1+E_2)}\Theta\left(1-\left|\frac{s-2m^2-2E_1E_2}{2p_1p_2}
\right|\right).
\end{eqnarray}

The $\Theta$ function constraint can be written as:
\[(E_1+E_2)^2m^2<sE_1E_2+m^2s-\frac{s^2}{4}.\]

Now we change integration variables from $E_1$, $E_2$ to $x=E_1E_2$, and
$y=E_1+E_2$. The collision rate becomes:
\begin{equation}
W=\frac{\pi\gamma^2}{2(2\pi)^4\mu^2}
\int_{4m^2}^\infty\mbox{d}s\;\sqrt{s(s-4m^2)}
\int\mbox{d}x\mbox{d}y\;\frac{\mbox{e}^{-y/T}}{\sqrt{y^2-4x}}
\Theta(x+m^2-\frac{s}{4}-\frac{y^2m^2}{s}).
\end{equation}

The integration over $x$ can be carried out first. The absolute lower bound of
$x=E_1E_2$ is $m^2$, but $x$ is also restricted by the $\Theta$ function. By
noticing $y\ge s$, and hence $y^2m^2/s-m^2+s/4>m^2$, we see the lower bound
should be $y^2m^2/s-m^2+s/4$. The upper bound is determined from the square
root in the integrand to be $y^2/4$. The result is:
\begin{equation}
\int_{y^2m^2/s-m^2+s/4}^{y^2/4}\mbox{d}x\;\frac{1}{\sqrt{y^2-4x}}=
\frac{1}{2}\sqrt{(s-4 m^2) (y^2/s-1)}.
\end{equation}

Now the collision rate per unit volume is reduced to a 2-dimensional integral:
\begin{equation}
W=\frac{\pi\gamma^2}{4(2\pi)^4\mu^2}
\int_{4m^2}^\infty\mbox{d}s\;(s-4m^2)
\int_{\sqrt{s}}^\infty\mbox{d}y\;\mbox{e}^{-y/T}\sqrt{y^2-s}.
\end{equation}

The $y$ integral can be readily carried out to be:
\begin{equation}
\int_{\sqrt{s}}^\infty\mbox{d}y\;\mbox{e}^{-y/T}\sqrt{y^2-s}=
\sqrt{s}TK_1\left(\frac{\sqrt{s}}{T}\right).
\end{equation}
In the above, $K_1(x)$ is the modified Bessel function.

Taking $\gamma=16$, we arrive at equation\ (\ref{rate3}).

\section{One example showing the interference of generations of distributions
of variables}

The following is the a sample program that generates the particle
momentum and position together. ran1 is taken from \cite{press1} and declared
as real*8 function.
\begin{verbatim}

      program dist
      implicit real*8 (a-h, o-z)
      parameter (mul = 40000)
      parameter (size = 5.19653401d0)
      external ran1
      common /para/ xmp
      temp = 0.5d0
      xmp = 5d0
      do i = 1, mul
         call energy(e, temp)
         call momentum(px, py, pz, e)
         x = 2d0 * ran1(iseed) - 1d0
         x = x * 5d0 * size
         y = 2d0 * ran1(iseed) - 1d0
         y = y * 5d0 * size
         z = 2d0 * ran1(iseed) - 1d0
         z = z * 5d0 * size
      end do
      stop
      end

      subroutine energy(e, temp)
      implicit real*8 (a-h, o-z)
      external ran1
      common /para/ xmp
 1000 continue
      e = ran1(iseed)
      e = e * ran1(iseed)
      e = e * ran1(iseed)
      if (e .le. 0d0) goto 1000
      e = - temp * log(e)
      if (ran1(iseed) .gt. 
     &     exp((e - sqrt(e ** 2 + xmp ** 2))/temp)) then
         goto 1000
      end if
      return
      end
        
      subroutine momentum(px, py, pz, e)
      implicit real*8 (a-h,o-z)
      external ran1
      parameter (pi = 3.14159265358979d0)
      cost = 2d0 * ran1(iseed) - 1d0
      sint = sqrt(1d0 - cost ** 2)
      phi = 2d0 * pi * ran1(iseed)
      px = e * sint * cos(phi)
      py = e * sint * sin(phi)
      pz = e * cost
      return
      end

\end{verbatim}

The momentum distribution is shown in Fig.~9 along with that generated by
ran3, and the theory prediction. We see that the momentum distribution is
correctly generated with reasonable fluctuations.

Fig.~10 gives the position $x$ distribution from the above program using ran1
and ran3. ran1 result has much larger fluctuations than the
expected statistical fluctuations, while ran3 result is consistent with the
expected statistical fluctuations. 

When the generation of momentum and position are separated, the distributions
all have reasonable fluctuations. Another possible (and more efficient) way to
solve the problem is to generate Gaussian momentum distribution for large $m/T$
values. 

Since with the above program, the total number of random number used is
$141931896$, while we can easily show the period of ran1 with the given
parameters is beyond $2000000000$, it is not clear that the large fluctuations
are due to the period of the random number generator ran1. But this is a
concrete example of ill-generated distribution.

{}

\newpage

\begin{table}
\caption{Input parameters to ZPC: cell size and 
screening mass for different temperatures and masses of
particles. $\mu_1$, $\mu_2$, $\mu_3$ correspond to three decreasing screening
masses as described in the text.\label{para1}}
\begin{center}
\begin{tabular}{c|c|c|c}
\hline\hline
   $T =     0.5$ GeV    &     m = 0 GeV  & m = 2.5 GeV  &  m = 5. GeV \\ \hline
   $l_{\mbox{\scriptsize{c}}}$ (fm)  
                        &    0.53240597  &  1.31503518  &  5.19653401 \\
   $\mu_1$ (fm$^{-1}$)  &    5.50178669  &  2.22745682  &  0.56368034 \\
   $\mu_2$ (fm$^{-1}$)  &    4.36677096  &  1.76793364  &  0.44739338 \\
   $\mu_3$ (fm$^{-1}$)  &    3.46590838  &  1.40320985  &  0.35509636 \\
\hline\hline
   $T =     1.0$ GeV    &     m = 0 GeV  & m = 2.5 GeV  &  m = 5. GeV \\ \hline
   $l_{\mbox{\scriptsize{c}}}$ (fm)  
                        &    0.26620298  &  0.36766446  &  0.65751757 \\
   $\mu_1$ (fm$^{-1}$)  &   11.00357361  &  7.96700354  &  4.45491373 \\
   $\mu_2$ (fm$^{-1}$)  &    8.73354210  &  6.32341486  &  3.53586735 \\
   $\mu_3$ (fm$^{-1}$)  &    6.93181691  &  5.01889767  &  2.80641975 \\
\hline\hline
   $T =     1.5$ GeV    &     m = 0 GeV  & m = 2.5 GeV  &  m = 5. GeV \\ \hline
   $l_{\mbox{\scriptsize{c}}}$ (fm)  
                        &    0.17746865  &  0.20981078  &  0.29320839 \\
   $\mu_1$ (fm$^{-1}$)  &   16.50536062  & 13.96107515  &  9.99011002 \\
   $\mu_2$ (fm$^{-1}$)  &   13.10031331  & 11.08091261  &  7.92915553 \\
   $\mu_3$ (fm$^{-1}$)  &   10.39772550  &  8.79492611  &  6.29337487 \\
\hline\hline
\end{tabular}
\end{center}
\end{table}

\newpage

\begin{table}
\caption{Measurement of non-causal collision number and total collision
collision number. 
\protect{\label{para2}}
}
\begin{center}
\begin{tabular}{c|c|c|c|c|c|c|c}
\hline\hline
   $\mu$ (fm$^{-1}$)    & a & l & $\chi_1$ & $\chi_2$ & 
   $N_{\protect{\mbox{\scriptsize{nonc}}}}$ &
   $N_{\protect{\mbox{\scriptsize{total}}}}$ & R \\ \hline
   5.50178669           &0.5&1  &0.5       &0.48      &3240             &
   31400           &10.3\%  \\
   4.36677096           &1  &1  &1         &0.88      &8250             &
   46100           &17.9\%  \\
   3.46590838           &2  &1  &2         &1.56      &19000            &
   64700           &29.4\%  \\
   3.46590838           &2  &5  &0.89      &0.81      &12500            &
   75400           &16.6\%  \\
   3.46590838           &2  &10 &0.63      &0.59      &9700             &
   77800           &12.5\%  \\ 
\hline \hline
\end{tabular}
\end{center}
\end{table}

\newpage

\begin{figure}
\hspace{0.5cm}
\psfig{figure=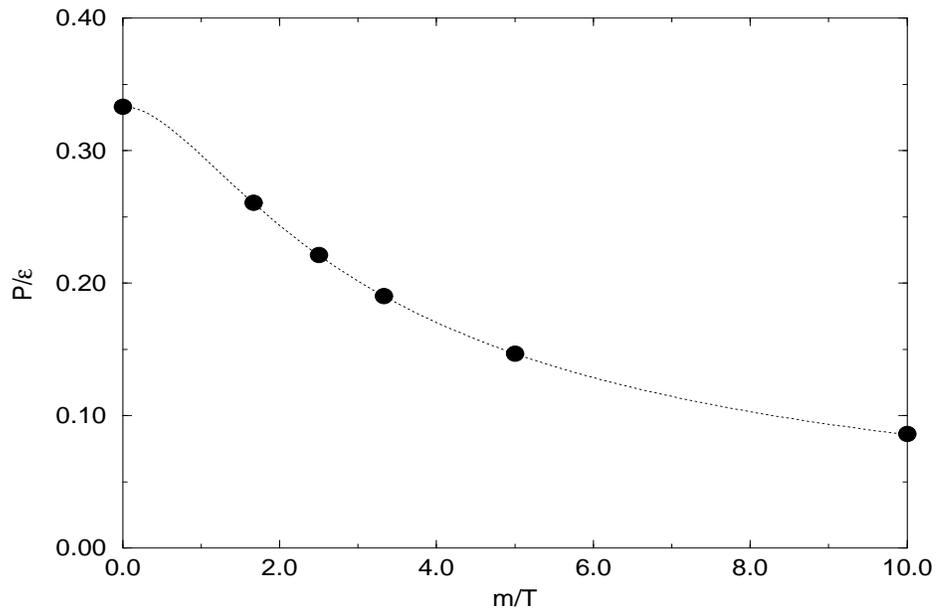,height=3.5in,width=3.0in,angle=-90} 
\caption{Pressure-energy density ratio $P/\epsilon$ as function of $m/T$. The
dotted curve is the analytic prediction, and the filled black dots are cascade
data.}
\end{figure}

\newpage

\begin{figure}
\hspace{0.5cm}
\psfig{figure=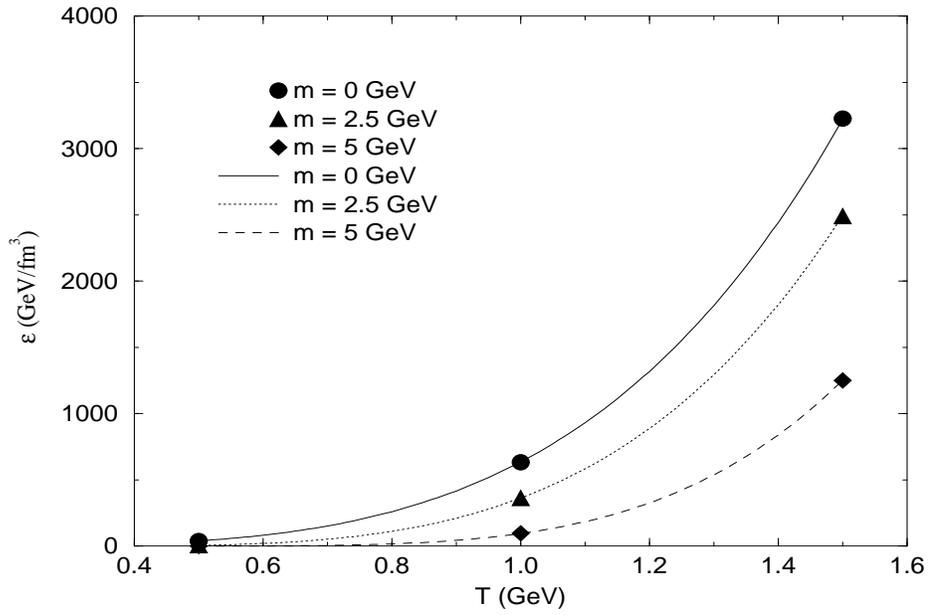,height=3.5in,width=3.0in,angle=-90} 
\caption{Energy density $\epsilon$ as a function of temperature $T$ for
different masses. The curves are predictions, and the dots are cascade data. }
\end{figure}

\newpage

\begin{figure}
\hspace{0.5cm}
\psfig{figure=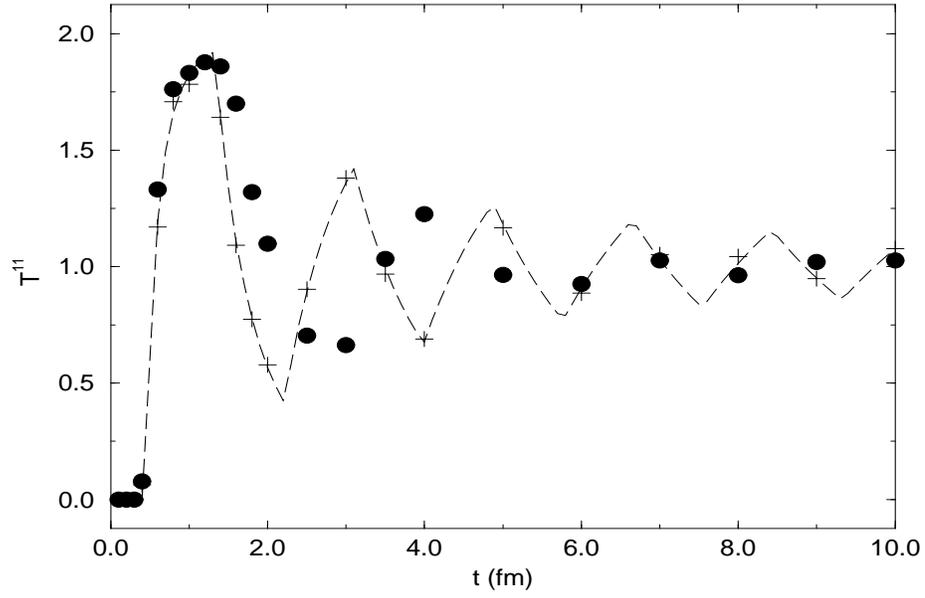,height=3.5in,width=3.0in,angle=-90} 
\caption{Evolution of the longitudinal momentum flux.
$T^{11}$ is in units of $T^{11}(\infty)$. 
%Initially all particles are confined to $0<x<L/2$. 
The size of the box is determined by the asymptotic temperature
$T=1.5$ GeV, and the screening mass is set to be $16.5$ fm$^{-1}$ 
for the interacting case.}
\end{figure}

\newpage

\begin{figure}
\hspace{0.5cm}
\psfig{figure=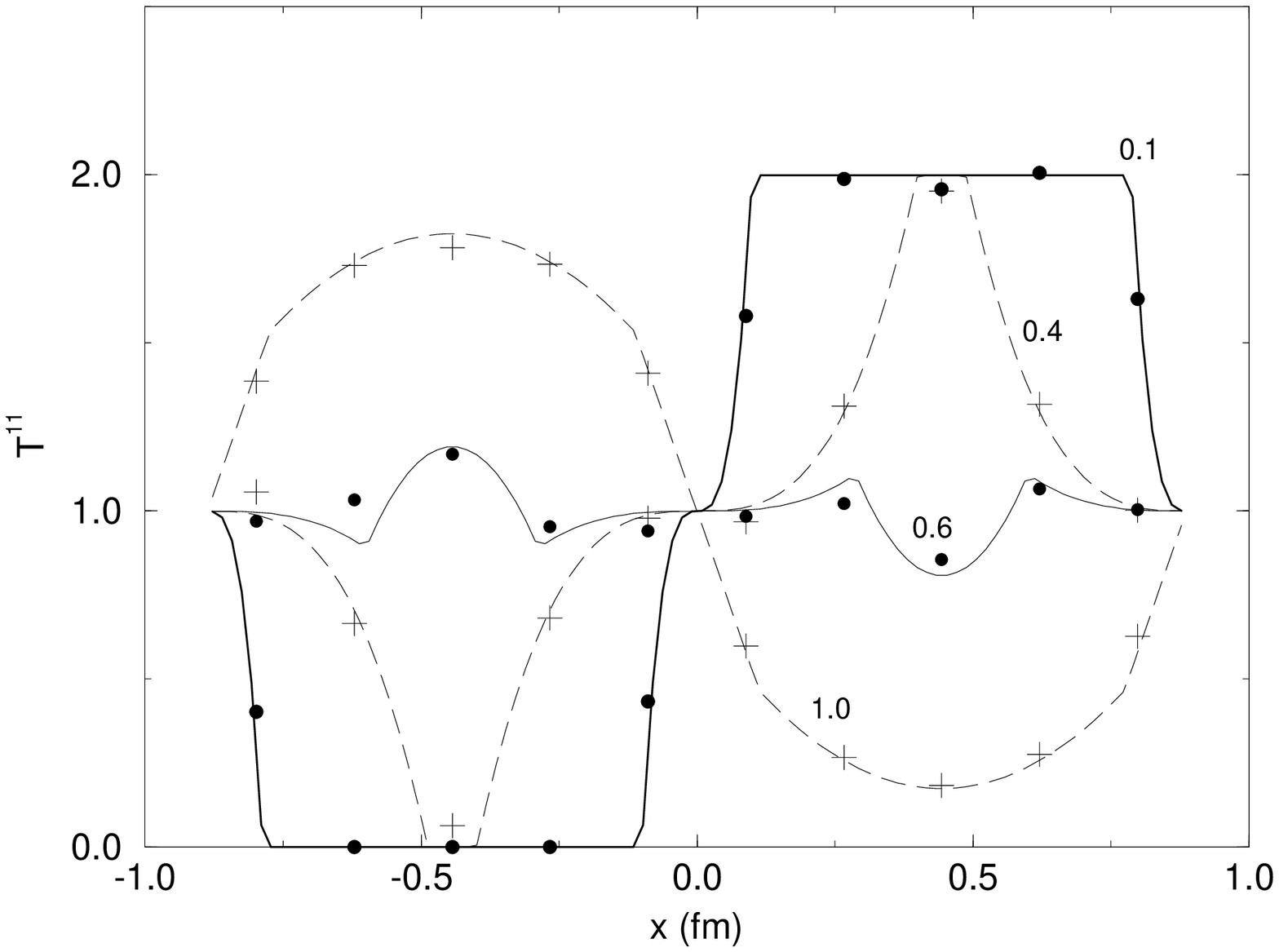,height=3.5in,width=3.0in,angle=-90} 
\caption{Evolution of $T^{11}$ spatial distribution. The predictions are drawn
with solid and dashed lines alternatively with numbers indicating the times in
fm. The cascade data are drawn in filled circles and pluses alternatively for
different times. Asymptotic temperature $T=1.5$ GeV.} 
\end{figure}

\newpage

\begin{figure}
\hspace{0.5cm}
\psfig{figure=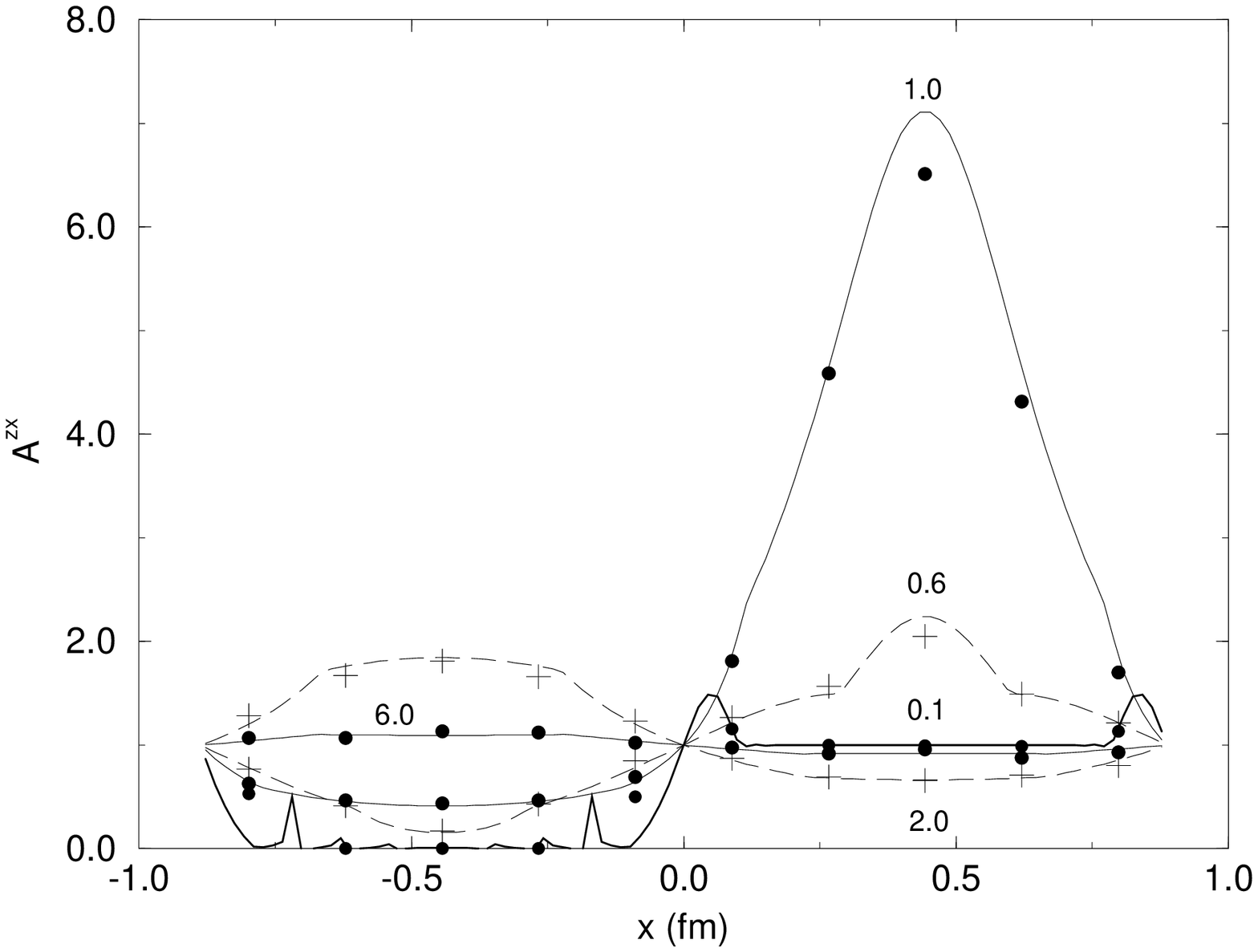,height=3.5in,width=3.0in,angle=-90} 
\caption{Evolution of $A^{zx}$ spatial distribution. The predictions are drawn
with solid and dashed lines alternatively with numbers indicating the times in
fm. The cascade data are drawn in filled circles and pluses alternatively for
different times. Asymptotic temperature $T=1.5$ GeV.} 
\end{figure}

\newpage

\begin{figure}
\hspace{0.5cm}
\psfig{figure=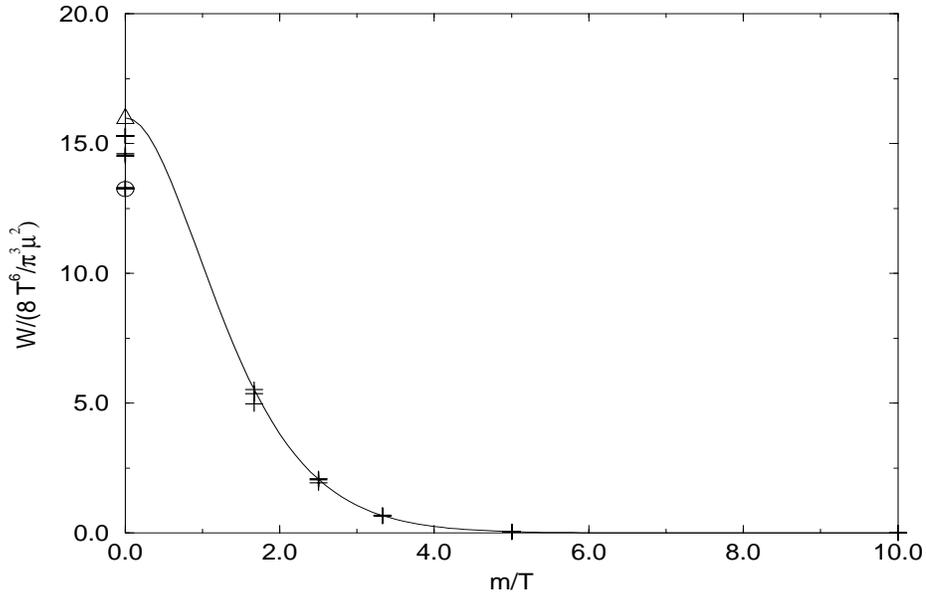,height=3.5in,width=3.0in,angle=-90} 
\caption{Collision rate per unit volume $W$ as a function of $m/T$. The curve
is prediction, and the pluses are cascade data. The open circle corresponds to
the rate at $m/T = 0$ with $\mu = 3.46590838$ fm$^{-1}$. 
The result from scaling the cross section down by a factor of $10$ and
increasing the particle density by a factor of $10$ is shown as the open 
triangle.}
\end{figure}

\newpage

\begin{figure}
\hspace{0.5cm}
\psfig{figure=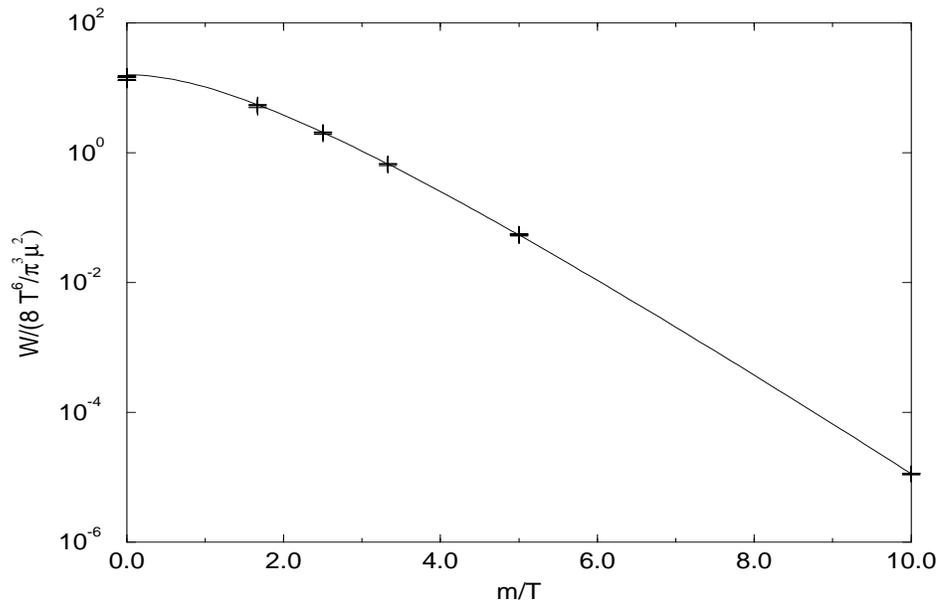,height=3.5in,width=3.0in,angle=-90} 
\caption{Collision rate per unit volume $W$ as a function of $m/T$. The curve
is prediction, and the pluses are cascade data.}
\end{figure}

\newpage

\begin{figure}
\hspace{0.5cm}
\psfig{figure=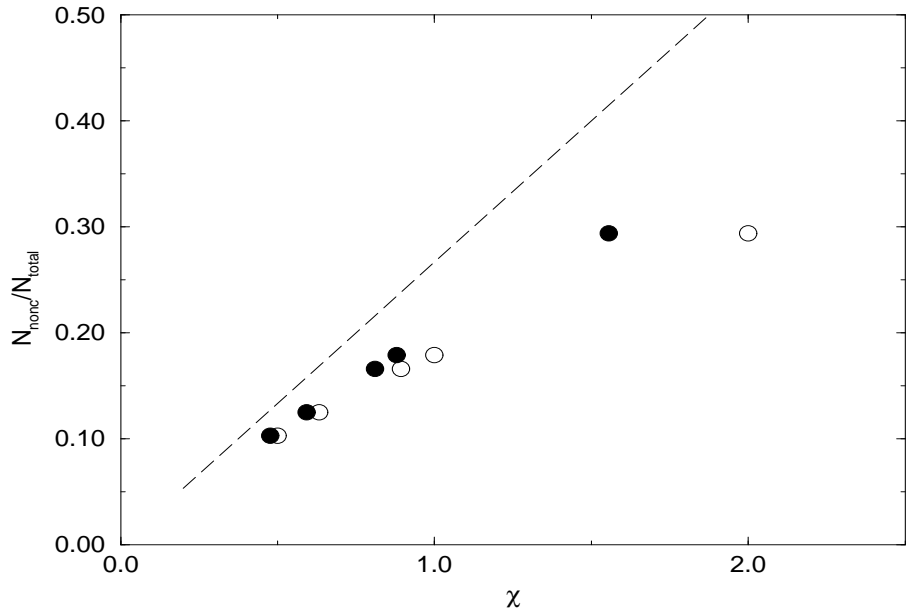,height=3.5in,width=3.0in,angle=-90} 
\caption{The number of non-causal collisions over the total number of
collisions as a function of the ratio of interaction length to the mean free
path. The dashed curve is the estimate when the radius of causal sphere is 
taken to be the interaction range. Open(filled) circles are data plotted
against $\chi_1$($\chi_2$) in table\ (\protect{\ref{para2}}).}
\end{figure}

\newpage

\begin{figure}
\hspace{0.5cm}
\psfig{figure=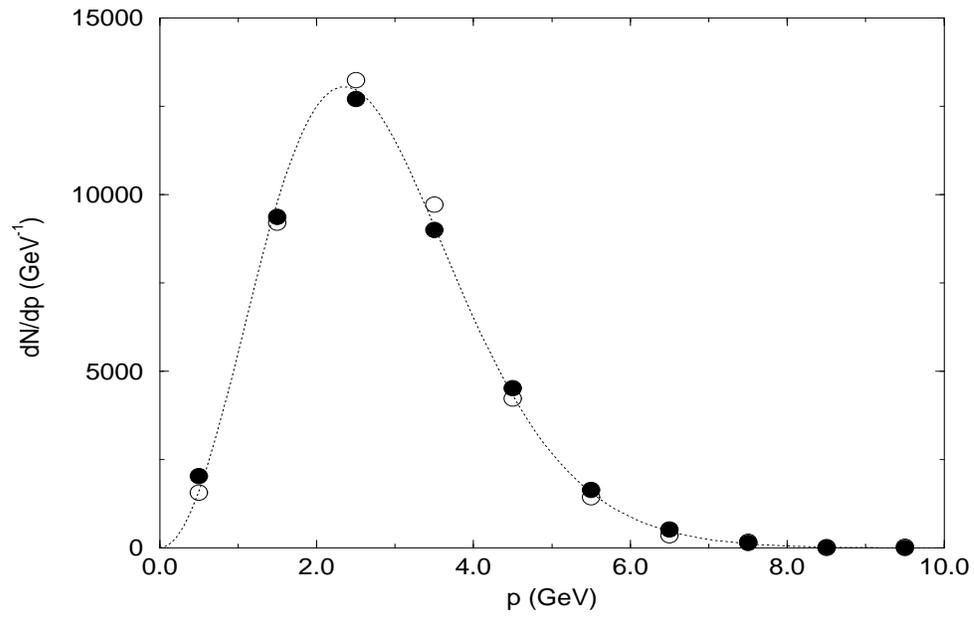,height=3.5in,width=3.0in,angle=-90} 
\caption{Momentum distribution for $m=5$ GeV, $T=0.5$ GeV with $40000$
particles. The dotted line is the prediction, the open circles come from code
using ran1, and the filled circles are ran3 results.} 
\end{figure}

\newpage

\begin{figure}
\hspace{0.5cm}
\psfig{figure=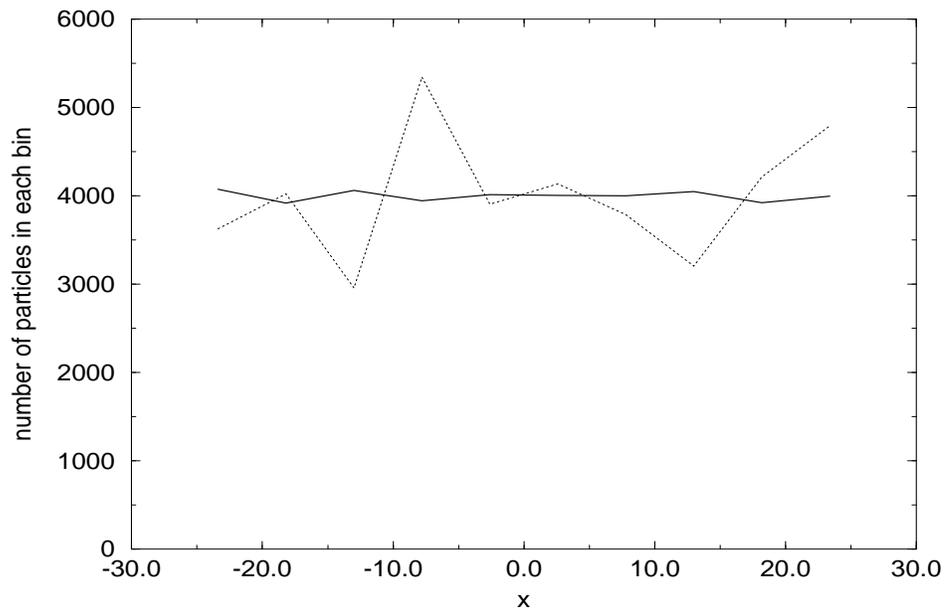,height=3.5in,width=3.0in,angle=-90} 
\caption{Number of particles per bin as a function of position $x$ for $m=5$
GeV, $T=0.5$ GeV with $40000$ particles . The dotted
curve is ran1 result, while the solid curve is for ran3.}
\end{figure}

\end{document}